**Modification of perpendicular magnetic anisotropy and domain wall velocity in Pt/Co/Pt by voltage-induced strain**

[1]P. M. Shepley, [2]A. W. Rushforth, [2]M. Wang, [1]G. Burnell, [1]T. A. Moore
[1]School of Physics and Astronomy, University of Leeds, Leeds, LS2 9JT, United Kingdom,
[2]School of Physics and Astronomy, University of Nottingham, Nottingham, NG7 2RD, United Kingdom.

Correspondence should be addressed to T. A. Moore (t.a.moore@leeds.ac.uk).

The perpendicular magnetic anisotropy $K_{eff}$, magnetization reversal, and field-driven domain wall velocity in the creep regime are modified in Pt/Co(0.85-1.0 nm)/Pt thin films by strain applied via piezoelectric transducers. $K_{eff}$, measured by the extraordinary Hall effect, is reduced by 10 kJ/m$^3$ by tensile strain out-of-plane $\varepsilon_z = 9 \times 10^{-4}$, independently of the film thickness, indicating a dominant volume contribution to the magnetostriction. The same strain reduces the coercive field by 2-4 Oe, and increases the domain wall velocity measured by wide-field Kerr microscopy by 30-100 %, with larger changes observed for thicker Co layers. We consider how strain-induced changes in the perpendicular magnetic anisotropy can modify the coercive field and domain wall velocity.

The study of magnetic domain wall motion in thin films and nanostructures with perpendicular magnetic anisotropy (PMA) is motivated by the desire to understand the fundamental physics at play and by the potential for applications in spintronic memory and logic [1-5]. The advantages of PMA materials are their stable magnetization states, narrow domain walls and promise of efficient current-induced domain wall motion [6]. The counterpoint to stability is a large energy barrier to magnetization reversal, necessitating large switching fields or currents. In the case of current-induced domain wall motion, a large PMA limits the threshold current density[7], determined by extrinsic pinning, to above $10^{11}$ A/m$^2$. Even with a decrease of an order of magnitude, the current required to drive the magnetization reversal and the consequent Joule heating would constrain the packing density of component nanostructures in memory devices, as well as waste energy [5]. There is thus much interest in reducing the energy barrier to magnetization reversal, for example by electric field [8-11] or mechanical strain [12-23]. Our approach is to use strain from piezoelectric transducers to modify the anisotropy in PMA materials and thus reduce the magnetic field needed for domain wall motion.

Strain-induced changes in magnetic anisotropy energy and hysteresis loops have been studied previously in hybrid piezoelectric/ferromagnet heterostructures where the magnetic layer has either in-plane [13,16,19] or perpendicular [12,14,15] magnetic anisotropy. Control of domain wall motion using the strain from a piezoelectric has been studied at room temperature in materials with in-plane anisotropy including FeGa thin films [16], CoFeB [18] and CoFe [17], and in (Ga,Mn)(As,P) with PMA at 90 K [21]. Large changes in domain wall velocities have been observed in glass-coated amorphous microwires under stress [22,23]. While there has been work on PMA materials where the domain wall velocity is modified by interface charging, rather than strain [8-11], to our knowledge there has as yet been no systematic experimental study of the effect of strain on domain wall motion in thin film PMA materials at room temperature.

Here we measure the change in PMA induced by strain in Pt/Co/Pt and study the consequent effects on magnetization reversal and field driven domain wall motion in the creep regime. Creep motion of magnetic domain walls in ultra-thin films with perpendicular magnetic anisotropy has been the focus of much attention in recent years [24-31]. Creep is a phenomenon that occurs in various physical systems when a one dimensional elastic interface is driven through a two dimensional weakly disordered landscape [24]. In a film of ultra-thin Pt/Co/Pt, magnetization reversal takes place by nucleation of very few reverse domains, with domain walls separating the reversed and unreversed regions. Applying a magnetic field H provides the driving force to move the domain walls and increase the size of the reversed regions. Below a critical field (the depinning field $H_{dep}$) the domain walls act as elastic strings that can become pinned by peaks in the magnetic anisotropy energy landscape of the film, described by the pinning energy barrier $U_c$. Fluctuations in thermal energy allow the domain walls to overcome the pinning barriers. The velocity of a magnetic domain wall is described by the creep law as

$$v = v_o \exp\left[-\frac{U_c}{kT}\left(\frac{H_{dep}}{H}\right)^{1/4}\right], \qquad 1$$

where kT is the product of Boltzmann's constant and temperature. The numerical prefactor $v_o$ is considered to be proportional to $l_{opt}$ – the lateral length of the small section of wall that undergoes a thermally assisted jump forwards prior to an avalanche [30,31].

RESULTS

**Magnetic anisotropy**

We begin by quantifying the change in PMA of Pt/Co/Pt under strain. Piezoelectric transducers were bonded to thin glass substrates onto which Ta(4.5nm)/Pt(2.5nm)/Co(t)/Pt(1.5nm) films had been sputter-deposited. We chose three Co thicknesses (t = 0.85nm, 0.95nm and 1.0nm) close to the reorientation transition from dominant perpendicular to in-plane magnetic anisotropy, which occurs at t = 1.1 nm in our films.

To strain a film, a voltage was applied to the piezoelectric transducer. The strain was measured from changes to the longitudinal resistance of the Hall bar devices patterned from the Pt/Co/Pt (see Supplementary Information). A positive voltage causes biaxial compression in the plane of the film that translates to a tensile out-of-plane strain up to a maximum of $\varepsilon_z$

= 9 x 10$^{-4}$ at 150 V. A voltage of -30 V gives a compressive out-of-plane strain of $\varepsilon_z$ = -3 x 10$^{-4}$.

To assess the effect of piezo-induced strain on the magnetic anisotropy energy of the Pt/Co/Pt films, measurements of anisotropy field were carried out by monitoring the extraordinary Hall effect (EHE) signal during magnetic field sweeps. The Hall resistance can be expressed as

$$R_H = (R_o H_z + R_H \mu_o m_z)/t .\qquad\qquad 2$$

The first term represents the ordinary Hall effect, where $R_o$ is the ordinary Hall coefficient. This effect is linear in applied out-of-plane field $H_z$ and is small enough in our measurements to be neglected in the analysis. The second term arises from the EHE, which is proportional to the out-of-plane magnetization $m_z$, with $\mu_o$ being the permeability of free space and $R_H$ the EHE coefficient. The size of the EHE resistance gives a measure of the component of the magnetization pointing out of the plane and can be used to determine PMA [32].

A current of 1 mA was passed along the Hall bar (x) and the Hall voltage monitored in an orthogonal in-plane direction (y) via one of the cross structures. A schematic of the measurement geometry is show in the inset to Figure 1a. To make a measurement, the plane of the device was first precisely aligned to an in-plane magnetic field by rotating the sample around the x axis until the Hall signal was as close to zero as possible during a field sweep along the y axis. An out-of-plane field was then applied to saturate the magnetization of the Pt/Co/Pt. Following this, an in-plane field was swept along the y axis from 0 Oe to 7000 Oe and the Hall resistance measured as the magnetization rotated from out-of-plane (maximum Hall signal) to in-plane (zero Hall signal). Figure 1a shows examples of the EHE data obtained. Initially (up to ~600 Oe in the case of Figure 1a), $m_z$ follows a parabola as expected if the magnetization were to rotate coherently (see Supplementary Information). As the field increases beyond 600 Oe, $m_z$ deviates from the parabola as the magnetization breaks up into domains with a size of ≤2 μm as measured by wide-field Kerr microscopy. The magnetization is eventually saturated in the plane, and the path $m_z$ would have followed if the magnetization had continued to rotate coherently is rejoined. The low field regime (up to ~600 Oe) where the moment rotates coherently is extrapolated, following the dashed lines in Figure 1a, to obtain the anisotropy field $H_k$, which is defined as the point where the extrapolated coherent rotation crosses $m_z = 0$. Proper alignment of the field to the plane of the device ensured that the films were truly saturated along an in-plane axis, allowing for direct comparison of $H_k$ between samples.

It can be seen from Figure 1a that applying a voltage to the transducer to strain the film results in a change in the anisotropy field. We relate the measured anisotropy fields to anisotropy constants $K_{eff}$ by

$$H_k = \frac{2}{\mu_0 M_s} K_{eff}. \qquad 3$$

In this expression, $M_s = (1.29 \pm 0.08) \times 10^6$ A/m is the saturation magnetization of Pt/Co/Pt, measured by SQUID VSM. The inset of Figure 1b shows the PMA constants $K_{eff}$ in the unstrained films measured by EHE. $K_{eff}$ decreases as the Co layer thickness increases, from $(210 \pm 10)$ kJ/m$^3$ for 0.85 nm and $(134 \pm 8)$ kJ/m$^3$ for 0.95 nm, to $(98 \pm 7)$ kJ/m$^3$ for 1.0 nm. The reduction in $K_{eff}$ indicates that the Co thickness consistently increases, and that the sub-nanometer precision of our thickness scale is valid. The anisotropy constants measured are consistent with 200 kJ/m$^3$ obtained previously for Pt/Co(1.0nm) multilayers [33] and with 400 kJ/m$^3$ obtained for Pt(4nm)/Co(0.5nm)/Pt(2nm) [34].

Figure 1b shows that tensile out-of-plane strain $\varepsilon_z$ reduces the PMA of the Pt/Co/Pt. The change per unit of strain is the same for the three Co thicknesses. We find a magnetostriction constant of $(-3.5 \pm 0.2) \times 10^{-5}$ from a least squares fit of the change in anisotropy to

$$\Delta K = \frac{3}{2} Y \lambda \varepsilon, \qquad 4$$

where Y is taken to be the average of the Young's moduli of bulk Co and Pt (180 GPa) [33], λ is the saturation magnetostriction and ε is the strain. A previous study of Pt/Co multilayers found a significant interface contribution to the magnetostriction [33]. Our measured magnetostriction constant is slightly lower than for bulk Co ($\lambda = -5 \times 10^{-5}$), is close to that of $Co_{90}Pt_{10}$ alloy [35], and does not change with the Co thickness. Since the magnetostriction of CoPt alloys increases from negative values at low Pt concentrations to positive values at high

Pt concentrations [35], the negative magnetostriction constant that we measure indicates that there is little intermixing at the Pt/Co interface. We conclude that in our samples the observed magnetostriction arises from the bulk Co volume.

**Magnetization reversal**

Next we study the effects of strain on the magnetization reversal of Pt/Co/Pt. Hysteresis loops were measured using polar Magneto-optical Kerr effect (MOKE). The sweep rate in the range of the coercive field was 2 Oe/s and five separate loops were averaged to obtain the data for each loop in Figure 2a. The polar MOKE hysteresis loops for 0.85, 0.95 and 1.0 nm Co all have the square shape typical of a perpendicular easy axis. The coercive field is largest for the 0.85 nm Co layer, which has the highest PMA, and decreases as the Co becomes thicker. For all three Co thicknesses the coercive field of the magnetic hysteresis loops is reduced by between 2 and 4 Oe under tensile strain (Figure 2b). As the PMA is modified, the energy barrier to magnetization reversal is lowered so that a smaller magnetic field is needed.

**Domain wall velocity**

Finally, we investigate the changes in magnetization reversal further by studying domain wall creep motion. We measure this using the wide-field Kerr microscopy technique described under Methods. Figure 3(a) shows the domain wall velocity v plotted against the applied driving field H and figure 3(b) shows the natural logarithm of v plotted against $H^{-1/4}$ for Pt/Co/Pt with 0.85, 0.95 and 1.0 nm of Co, with the piezoelectric transducer at 0V (unstrained) and 150V (tensile strain). The linear behaviour of all datasets in Figure 3b and the fitting of Equation 1 is consistent with domain wall motion in the creep regime. Under tensile strain the velocity of the domain walls increases and we observe that the difference in domain wall velocity between 0 V and 150 V increases with applied field. Figure 4 shows the ratio of the domain wall velocity under tensile strain to the velocity in the unstrained state (0V). For t = 1 nm Co we observe a strain induced increase in the domain wall velocity by a factor of 2 measured at a magnetic field of 108 Oe, corresponding to an unstrained domain wall velocity of 60 µm/s.

The change in domain wall velocity with strain increases with Co thickness and is largest for t = 1.0 nm Co, so that it was possible to resolve changes at lower transducer voltages in this Hall bar. Figure 3c shows the natural logarithm of v with the transducer voltage at -30 V, 50 V and 100 V in addition to the 0 V and 150 V data shown in Figure 3b. For a given field, the velocity increases with increasing tensile strain (increasing positive voltage on the transducer), which corresponds to decreasing PMA.

The creep law (Equation 1) was fitted with a least squares method to the data in Figures 3b and 3c. The intercept of the fit with the vertical axis is $\ln v_o$ and the gradient of the line is the product $H_{dep}^{1/4} U_c/kT$. Figures 5a and 5b show how $\ln v_o$ and $H_{dep}^{1/4} U_c/kT$ vary with Co thickness in the strained and unstrained Pt/Co/Pt. Since we do not have a direct measure of $H_{dep}$, $U_c/kT$ is extracted by assuming that $H_{dep} = H_c$, which is reasonable because it accommodates the change with strain of $H_{dep}$ (proportional to the change in $H_c$). A comparison of the values of $H_c$ (Figure 2b) to the range of applied fields driving domain wall velocity (Figure 3a) shows this estimate of $H_{dep}$ to be too low; $H_c$ is within the range of fields that drive creep motion. The estimate of $H_{dep}$ produces an increased $U_c/kT$, but allows for a shift in $H_{dep}$ under strain equal to the shift in $H_c$. Figure 5c shows the measured values of $U_c/kT$. At 0 V it is found to be 69 ± 2 for 0.85 nm, 87 ± 2 for 0.95 nm and 87 ± 1 for 1.0 nm. As these values are artificially inflated by the estimate of $H_{dep}$, they are somewhat larger than values found in similar polycrystalline Pt/Co/Pt films [26], and to epitaxial Pt/Co/Pt films [27].

The values of the parameters shown in Figure 5 are lower in the t = 0.85 nm sample, while the values in the two thicker films are similar. They depend strongly on the microstructure of the material (such as size of crystal grains and film roughness), which may not be due to sample thickness alone. Any changes in $\ln v_o$ and $H_{dep}^{1/4} U_c/kT$ with strain are small compared with the uncertainties in the values, and unlike the shift in the domain wall velocity, no systematic trends can be observed. The precision of the measurements is limited by the low range of magnetic field we can measure over. As the applied field increases, the nucleation density increases, so that distance of travel of a domain wall before domains coalesce is reduced, limiting the measurable distance.

DISCUSSION

We find that under tensile out-of-plane strain $\varepsilon_z = 9 \times 10^{-4}$ the PMA of Pt/Co/Pt is reduced by 10 kJ/m$^3$, and the domain wall creep velocity is increased by up to a factor of 2, depending on the Co thickness. The coercive field is reduced under tensile strain, which may be the result of two effects: the smaller nucleation field as seen in the hysteresis loops, and the faster domain wall motion under strain, both of which may be linked to the perpendicular anisotropy. The experimental uncertainties arising from the limited magnetic field range in our measurements makes it difficult to exclude the possibility of any change in the pinning energy $U_c$ as a function of strain. We note that the pinning energy is specific to domain wall creep and thus not necessarily a good indicator of coercivity.

We now consider how the modification of the anisotropy energy will affect the structure and energy of a domain wall. As the tensile strain increases and the PMA is reduced, the Bloch domain wall width $\delta = \pi\sqrt{A/K_{eff}}$ (where A is the exchange stiffness and $K_{eff}$ is the magnetic anisotropy constant) increases and the domain wall energy $\gamma = 4\sqrt{AK_{eff}}$ decreases. Therefore, reduced PMA lowers the energy barrier to elements of the film reversing as the domain wall moves, so for lower $K_{eff}$ one might expect a lower nucleation field and a larger domain wall velocity for a given driving field. The change in domain wall energy is proportionally larger when $K_{eff}$ is a relatively low value, which may explain the observed trend of larger velocity changes for thicker films, but is not sufficient to fully account for the doubling of velocity in the thickest sample. In creep motion, thermally assisted jumps can result in an increase in energy as the wall lengthens to encompass a newly reversed region. A larger reduction in domain wall energy with strain, as for the thicker films, might thus be expected to lead to a greater decrease in the energy required to move a section of wall forward, contributing to the observed larger change in velocity for thicker Co.

In summary, the PMA of Pt/Co/Pt has been modified by strain induced by a piezoelectric transducer. Lowering the PMA with strain reduces the coercive field of the Pt/Co/Pt and increases the domain wall creep velocity by up to 100% in the field range accessible in our experiments.

## METHODS

Ta(4.5nm)/Pt(2.5nm)/Co(t)/Pt(1.5nm) multilayers were deposited by dc magnetron sputtering onto 150µm thick glass substrates at room temperature. The thicknesses of the metallic layers were estimated using sputtering rates found from X-ray reflectivity of Co/Pt multilayers and Ta films. Glass microscope cover slips were chosen because they are thin enough to permit an isotropic transmission of the majority of the strain generated by the transducer. The base pressure of the sputtering system was $3 \times 10^{-8}$ Torr and the Ar pressure during sputtering was $2.4 \times 10^{-3}$ Torr. The multilayers were patterned into 50 µm wide Hall bars using standard optical lithography. The glass substrates with Hall bar devices on top were then bonded with epoxy resin to piezoelectric transducers (commercially available from Piezomechanik GmbH).

The domain wall velocity was measured using wide field Kerr microscopy. A reverse domain is nucleated, either in the Hall bar or in a small region of sheet film, with a short magnetic field pulse. Nucleation occurs at a few sites and the domains expand so that an approximately straight domain wall moves into the field of view of the microscope [25]. An image is recorded, then another magnetic field pulse is applied to move the domain wall. Another image is recorded and the difference between the two images is used to extract the distance the domain wall travels. This is divided by the length of the magnetic field pulse and the resulting velocity can be plotted against driving field as in Figure 3a. The lengths of the field pulses, defined at the full-width-half-maximum, were between 200 ms and 20 s, with a rise time of no more than 100 ms.

## ACKNOWLEDGEMENTS

The authors acknowledge financial support from EPSRC (Grant No. EP/K003127/1) and EU ERC Advanced Grant 268066. AWR acknowledges support from a Career Acceleration Fellowship (Grant No. EP/H003487/1), EPSRC, UK.

## AUTHOR CONTRIBUTIONS

P.M.S. deposited the Pt/Co/Pt layers and carried out the measurements. M.W. and A.W.R. contributed to device fabrication. P.M.S., A.W.R., M.W., G.B. and T.A.M. contributed to the design of the experiments, analysis of the results and preparation of the manuscript.

ADDITIONAL INFORMATION

There authors have no competing financial interests to declare.

FIGURES

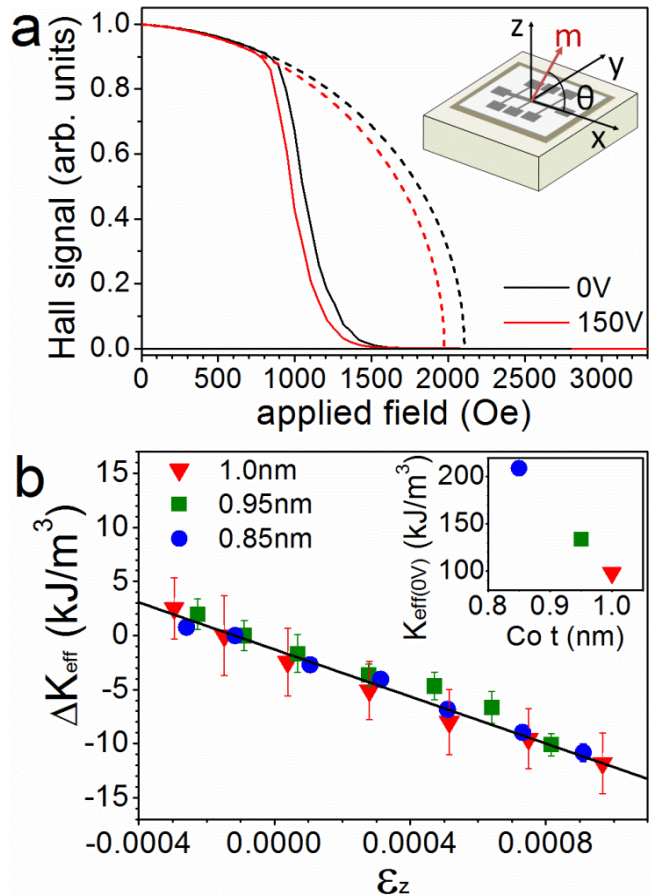

Figure 1 a) Examples of normalised EHE data from Pt/Co(0.95 nm)/Pt used for finding the anisotropy field $H_k$. The solid lines show the normalised data with 0 V (black lines) and 150 V (orange lines) applied to the transducer. The dashed lines are extrapolated from fits to the data below 600 Oe. $H_k$ is the applied field at which the extrapolated curves meet the line where the Hall signal is 0. The inset is a schematic of a Hall bar on a transducer showing the measurement geometry. b) The change in the PMA constant $K_{eff}$ of Pt/Co(t)/Pt (t = 0.85, 0.95, 1.0 nm) due to out-of-plane strain $\varepsilon_z$ induced by piezoelectric transducers. The solid line is a fit of the data to Equation 4. The inset gives the anisotropy constants of the three unstrained films against Co thickness. The error bars in the inset are smaller than the data points.

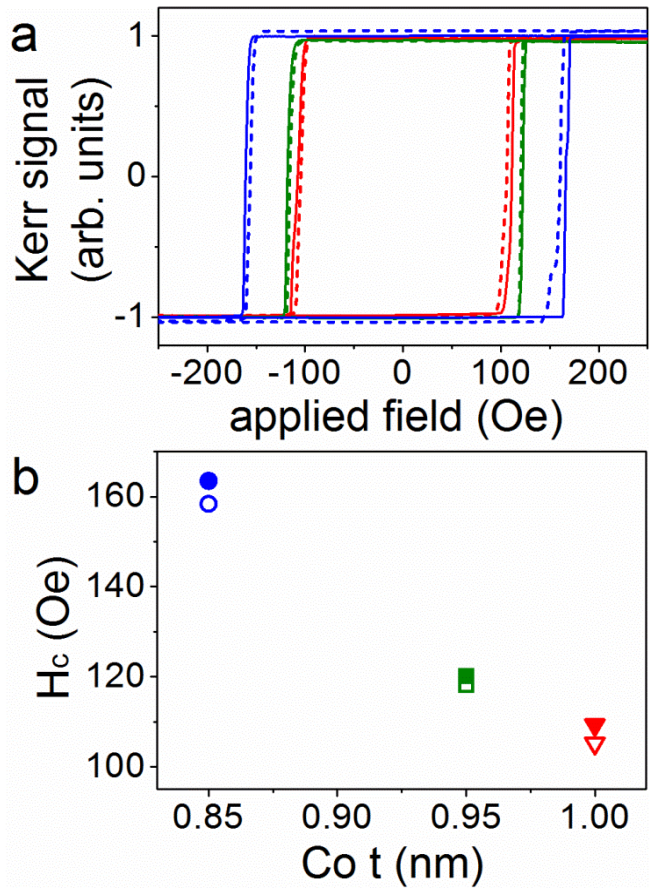

Figure 2 a) Polar MOKE hysteresis loops of Pt/Co(t)/Pt with t = 0.85 nm (blue lines), 0.95 nm (green lines) and 1.0 nm (red lines). The solid lines represent the unstrained films and the dashed lines show the hysteresis loops under tensile out-of-plane strain induced by applying 150 V to the piezoelectric transducers. b) The coercive fields of the hysteresis loops are plotted against Co thickness. The solid shapes are the unstrained films and the open shapes are the strained films. The error bars are smaller than the data points.

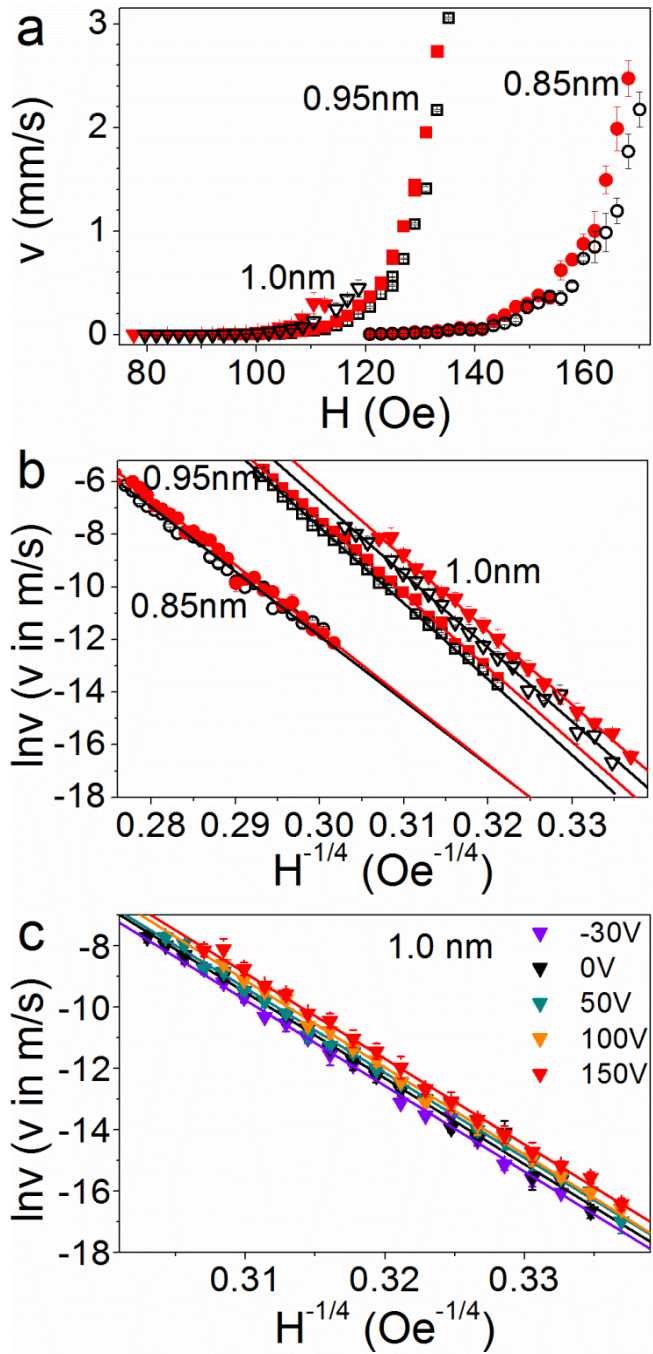

Figure 3 a) Domain wall velocity v plotted against applied field H and b) natural logarithm of v plotted against $H^{-1/4}$. Both plots show data for unstrained Pt/Co(t)/Pt (black open shapes) and Pt/Co(t)/Pt under tensile out-of-plane strain induced by applying 150 V to the piezoelectric transducers (red solid shapes), for t = 0.85 (circles), 0.95 (squares) and 1.0 nm (triangles). c) natural logarithm of v plotted against $H^{-1/4}$ for Pt/Co(t)/Pt t = 1.0 nm with the transducer at voltages of -30, 0, 50, 100 and 150 V. The straight lines in b and c are fits of Equation 1 to the data.

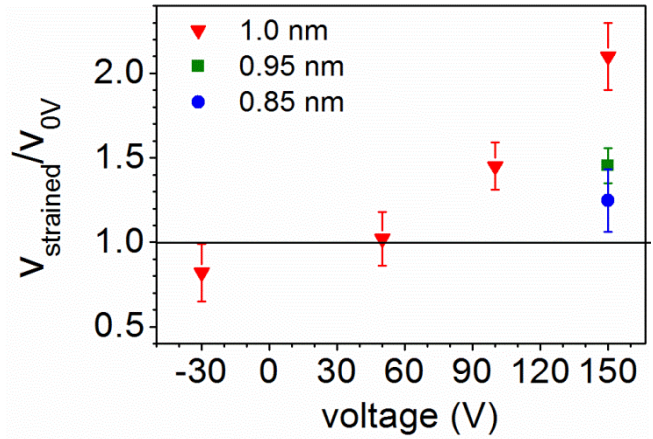

Figure 4 The ratio of domain wall velocity in the unstrained Pt/Co(t)/Pt to the velocity in strained Pt/Co(t)/Pt plotted against transducer voltage for t = 1.0 nm, 0.95 nm and 0.85 nm. The unstrained velocity in each film is ~60 µm/s and the line at $v_{strained}/v_{0V} = 1$ represents $v = v_{0V}$.

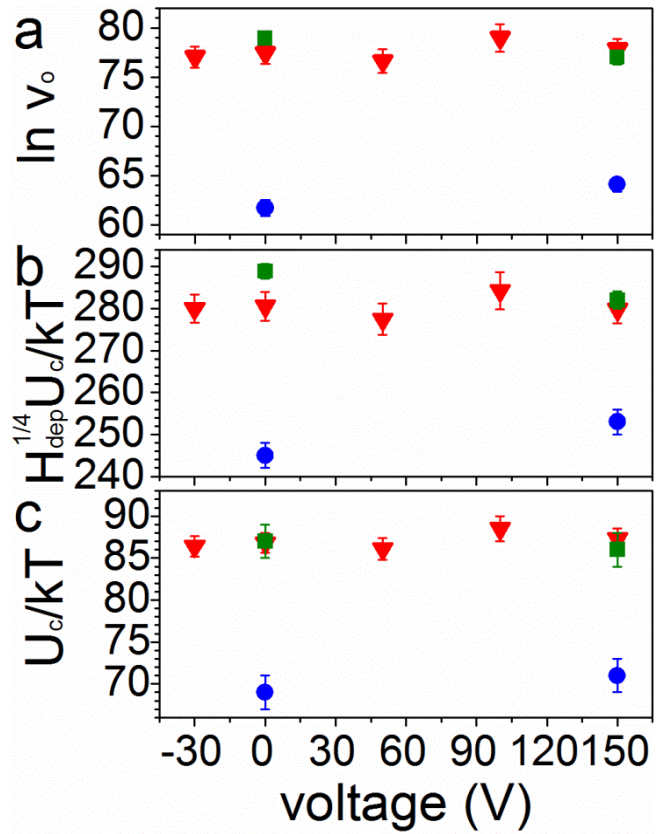

Figure 5 a) The values of the intercepts with the ln v axis extracted from fits of the creep law (Equation 1) to the data in Figures 3b and 3c plotted against the voltage applied to the transducers. b) Gradients of the fits of the creep law to data in Figure 3b and 3c plotted against the voltage applied to the transducers. c) Ratio of the pinning energy to thermal energy kT obtained by assuming $H_{dep} = H_c$. Red triangles represent Pt/Co(t)/Pt t = 1nm, green squares are t = 0.95 nm and blue circles are t = 0.85 nm.

16